\documentclass[fleqn,usenatbib,useAMS]{mnras}
\usepackage{newtxtext,newtxmath}
 
\usepackage[flushleft]{threeparttable}
\usepackage[caption=false]{subfig}
\usepackage{lscape}
\usepackage[T1]{fontenc}
\DeclareRobustCommand{\VAN}[3]{#2}
\let\VANthebibliography\thebibliography
\def\thebibliography{\DeclareRobustCommand{\VAN}[3]{##3}\VANthebibliography}
\hypersetup{linkcolor=blue,citecolor=blue,filecolor=cyan,urlcolor=magenta}
\usepackage{graphicx}	 
\usepackage{amsmath}	 
\usepackage{amssymb} 
\usepackage{float}
\usepackage{multirow}
\title[Nebular Attenuation Curve]{Variation of the Nebular Dust Attenuation Curve with the Properties of Local Star-forming Galaxies}

\author[S. Rezaee et al.]{
Saeed Rezaee$^{1}$\thanks{E-mail: saeed.rezaee@email.ucr.edu},
Naveen Reddy$^{1}$,
Irene Shivaei$^{2,3}$,
Tara Fetherolf$^{1}$,
Najmeh Emami$^{4}$,
A. A. Khostovan$^{5,6}$
\\
$^{1}$Department of Physics and Astronomy, University of California Riverside, Riverside, CA 92521, USA\\
$^{2}$Steward Observatory, University of Arizona, 933 North Cherry Avenue, Tucson, AZ 85721, USA\\
$^{3}$Hubble Fellow\\
$^{4}$Minnesota Institute for Astrophysics, University of Minnesota, 116 Church St SE, Minneapolis, MN 55455, USA\\
$^{5}$Astrophysics Division, NASA Goddard Space Flight Center, Greenbelt, MD 20771, USA\\
$^{6}$NASA Postdoctoral Program Fellow
}

\date{Accepted XXX. Received YYY; in original form ZZZ}

\pubyear{2021}

\newcommand{\ebmvgas}{E(B-V)_{\rm neb}}

\newcommand{\ha}{\text{H$\alpha$}}
\newcommand{\hb}{\text{H$\beta$}}
\newcommand{\hg}{\text{H$\gamma$}}
\newcommand{\hd}{\text{H$\delta$}}
\newcommand{\he}{\text{H$\epsilon$}}

\newcommand{\angstrom}{\text{\normalfont\AA}}
\begin{document}
\label{firstpage}
\pagerange{\pageref{firstpage}--\pageref{lastpage}}
\maketitle

\begin{abstract}
We use a sample of $78,340$ star-forming galaxies at $z\simeq 0.04-0.1$ from the SDSS DR8 survey to calculate the average nebular dust attenuation curve and its variation with the physical properties of galaxies. Using the first four low-order Balmer emission lines ($\ha,\hb,\hg,\hd$) detected in the composite spectrum of all galaxies in the sample, we derive a nebular attenuation curve in the range of $0.41\,\mu$m to $0.66\,\mu$m that has a similar shape and normalization to that of the Galactic extinction curve (Milky Way curve), the SMC curve and the nebular attenuation curve derived recently for typical star-forming galaxies at $z\sim2$. We divide the galaxies into bins of stellar mass, gas-phase metallicity, and specific star-formation rate, and derive the nebular attenuation curve in each of these bins.  This analysis indicates that there is very little variation in the shape of the nebular dust attenuation curve with the properties used to bin the galaxies, and suggests a near universal shape of the nebular dust attenuation curve at least among the galaxies and the range of properties considered in our sample.  

\end{abstract}
\section{Introduction}
\defcitealias{Reddy_2020}{R20}
\label{sec:intro}

Many of the key inferred physical properties of galaxies are sensitive to the effects of dust. For instance, the use of the unobscured rest-frame UV light from massive young stars or the nebular emission lines to estimate star formation rate (SFR) must be accompanied by a proper dust correction to account for the light absorbed and re-radiated by dust (e.g., \citealt{Kennicutt2009,Hao2011,Kennicutt2012}).
In general, the dust corrections applied to the stellar continuum may differ from those applied to nebular lines because the sightlines to H{\sc ii} regions may have a different distribution of dust (or dust with different properties) compared to sightlines towards non-ionizing stellar populations. \citep{Calzetti94,Charlot}. Nebular regions may contain dust grains with different size and mass properties \citep{Draine_2003} because of the presence of the strong radiation fields around massive stars \citep{Martinez,Hoang}. In addition, many studies have found a larger reddening for nebular emission lines versus the stellar continuum (e.g., \citealt{ Fanelli,Calzetti97,Calzetti00,Forster,Yoshikawa,Wild11,Wuyts,Kreckel,Kashino,Wuyts_13,Price,Reddy_2015,DeBarros,Buat,Koyama,Shivaei2020}). Thus, knowledge of the dust geometry and properties in different regions within galaxies is crucial for identifying and applying the appropriate dust corrections. The dust extinction/attenuation curves provide invaluable information on dust properties and dust distribution \citep{Draine&li}.


Extinction curves have been studied for the Milky Way (MW) and nearby galaxies, such as the Large and Small Magellanic Clouds and M31, by measuring the extinction along individual sightlines (e.g., \citealt{Nandy_80,Nandy75,Rocco,Bianch,Claytonm31}). The average total extinction curves for these galaxies are determined by combining these individual sightlines \citep{Seaton,Prevot,Cardelli89,Pei,Gordon03,fitzpatric}. There are major differences between the extinction curves derived for different sightlines within a galaxy and also the average curves for different galaxies. For example, \citet{Fitzpatrick90} showed a broad range of extinction curves for various Milky Way sight lines. In addition, comparing the average curves derived for the Milky Way \citep{Cardelli89}, Magellanic clouds  \citep{fitzpatric,Gordon03}, and M$31$ \citep{Bianch}  shows variations in both  UV/optical slope and strength of the UV bump (a broad extinction feature of the curve near $2175$ \AA). For external galaxies, extinction curves cannot be directly measured due to limited spatial resolution.  Nevertheless, one can compute attenuation curves that reflect the average wavelength dependence of dust obscuration and which depend on both the properties of the dust and the geometry of that dust with respect to the stars \citep{Charlot,Calzetti2001,Weingartner,Li2001,Conroy2010,conrey2013,Cheva,Krieck&cor,Reddy_2015,Shivaei2020,Buat2011,Buat2012}.  A wide range of attenuation curves that apply to the stellar continuum have been derived with different UV bump strengths and optical/UV slopes \citep{Calzetti00,Conroyb,Cheva,Reddy_2015,Salim18}. Many of these same studies, as well as others, have suggested that these variations in the stellar attenuation curve may be correlated with certain properties of galaxies, including their stellar mass, SFR, and metallicity (e.g., for low-redshift galaxies: \citealt{Johnson}, \citealt{Wild11}, \citealt{Battisti}, \citealt{Battisti17}, and for high-redshift samples: \citealt{Krieck&cor} , \citealt{Zeimann}, \citealt{Reddy_2015},  \citealt{Salmon}, \citealt{Shivaei2020}). In parallel, theoretical work has explored the variation in curves due to dust-star geometry and age \citep{Witt,Weingartner,Narayanan}. 

On the other hand, despite very recent work in quantifying the shape of the nebular dust attenuation curve at high redshift~(\citealt{Reddy_2020}, hereafter refer to as \citetalias{Reddy_2020}), there is little information on how the shape of the nebular curve may vary from galaxy-to-galaxy and with galaxy properties. The shape of the nebular dust curve is critical to inferring several important physical parameters of the ISM including gas-phase metallicity, ionization parameters, and star-formation rate derived from Balmer lines. The MW curve \citep{Cardelli89} is preferred to correct the nebular lines for the dust extinction as it is derived based on the sightline measurements of nebular regions \citep{Calzetti94,Wild_a,Liu,salimreview}. Additionally, \citetalias{Reddy_2020} found that the nebular attenuation curve for high-redshift galaxies is similar to that of the MW at rest-frame optical wavelengths.  However, the small sample size in that work prevented a detailed study of how the nebular dust attenuation curve varies with galaxy properties. To better understand the conditions that may shape the nebular attenuation curve, we take advantage of a large sample of local star-forming galaxies for which the nebular attenuation curve can be inferred.   

In this paper, we derive the nebular attenuation curve for local star-forming galaxies and examine its variation with stellar mass, specific SFR (sSFR), and gas-phase abundances, with the goal of understanding how these properties may influence the shape of the nebular attenuation curve, and hence dust properties and geometry, as a function of these properties. The initial work of \citetalias{Reddy_2020} laid the foundation for deriving the nebular attenuation curve for high-redshift galaxies. Here, we expand upon this work by examining the variation of the curve with stellar mass, sSFR, and oxygen abundance using a large sample of local star-forming galaxies drawn from the SDSS. The large sample size allows us to group the galaxies by various properties and still retain a sufficient number of galaxies in each bin to robustly derive the nebular attenuation curve.

The structure of this paper is as follows. In Section~\ref{sec:sample}, we outline the sample used in this work. Section~\ref{sec:composite} presents the approach to constructing composite spectra. Section~\ref{sec:calculations} describes the method used to derive the shape of the nebular attenuation curve. Section~\ref{sec:subsample} discusses the comparison between the nebular attenuation curves derived for each subsample in stellar mass, metallicity, and sSFR. Section~\ref{sec:discuss} presents  a discussion of the variation of the curve with the aforementioned properties.
We adopt a cosmology with $H_{0}=70$\,km\,s$^{-1}$\,Mpc$^{-1}$, $\Omega_{\Lambda}=0.7$, and $\Omega_{\rm m}=0.3$. All wavelengths are presented in the vacuum frame.

\begin{figure*}
    \centering
    \includegraphics[width=1\textwidth]{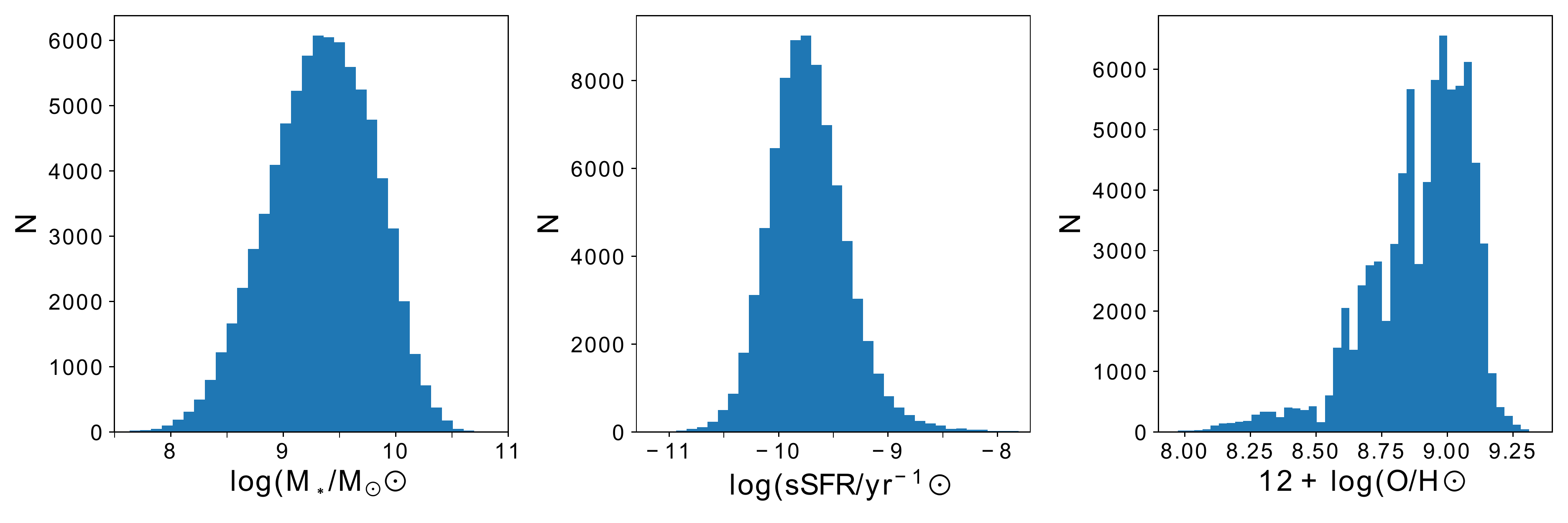}
    \caption{ Distribution of stellar mass (left), sSFR(middle), and gas-phase metallicity (right) of the sample analyzed in this work, and includes a total of $78,340$ low-redshift star-forming galaxies from the SDSS. }
    \label{fig:properties}
\end{figure*}

\section{\bf{sample}} 
\label{sec:sample}
In this study, we use optical spectroscopic observations of galaxies from the Sloan Digital Sky Survey Data Release 8 \citep{Aihara_2011}. Our sample is constructed using the publicly-available $Galspec$ catalogs provided by the MPA/JHU group~\citep{Kauffmann_2003,Brinchmann_2004,Tremonti2004}, and includes $78,340$ galaxies, all meeting the following criteria:
\begin{itemize}
    \item \textbf {(i) Only star-forming galaxies}: galaxies that lie below the active galactic nucleus (AGN) demarcation line of  \citet{Kauffmann_2003}.    
    \item \textbf{(ii) A redshift range of} $0.04\leq z \leq 0.1$: to ensure that the portion of galaxy which is measured inside the fiber aperture is reasonably representative of the entire galaxy. \\
\end{itemize}

The $Galspec$ catalogs include emission line measurements and inferences of galaxy properties. We refer the reader to \citet{Aihara_2011} for further details. In brief, line fluxes are corrected for the effect of stellar absorption using \citet{Bruzual_Charlott2003} stellar population synthesis models. The measurements of individual galaxy properties correspond to those obtained for the $3''$ SDSS fiber, and include stellar mass, sSFR, and gas-phase abundances. Stellar masses are based on fitting stellar population models to $ugriz$ photometry, and assume a  \citet{Kroupa_2001} initial mass function. Gas-phase abundance ($\mathrm{12+\log(O/H)}$), hereafter referred to
as the metallicity,  are calculated from the strong optical emission lines ($\rm [\ion{O}{II}]\lambda3727$, $\hb$, $\rm [\ion{O}{III}]\lambda5007$, $\rm [\ion{N}{II}]\lambda6548$, $6584$, and $\rm [\ion{S}{II}]\lambda6717$, $6731$) using the Bayesian methodology from \citet{Tremonti2004}, and \citet{Brinchmann_2004}. Star-formation rates are based on dust-corrected $\ha$ emission as described in \citet{Brinchmann_2004}. The sample used in this work spans the following range in physical properties: $6.68 < \log(M_{\ast}/M_{\odot}) < 11.46 $, $-11.79 < \log(sSFR/\rm yr^{-1}) < -7.05 \mathrm{, and}$ ${7.85 < 12+\log (\rm O/\rm H) < 9.40}$. Figure~\ref{fig:properties} shows the distribution of the physical properties of galaxies in this sample.

\begin{figure}
	\includegraphics[width=\columnwidth]{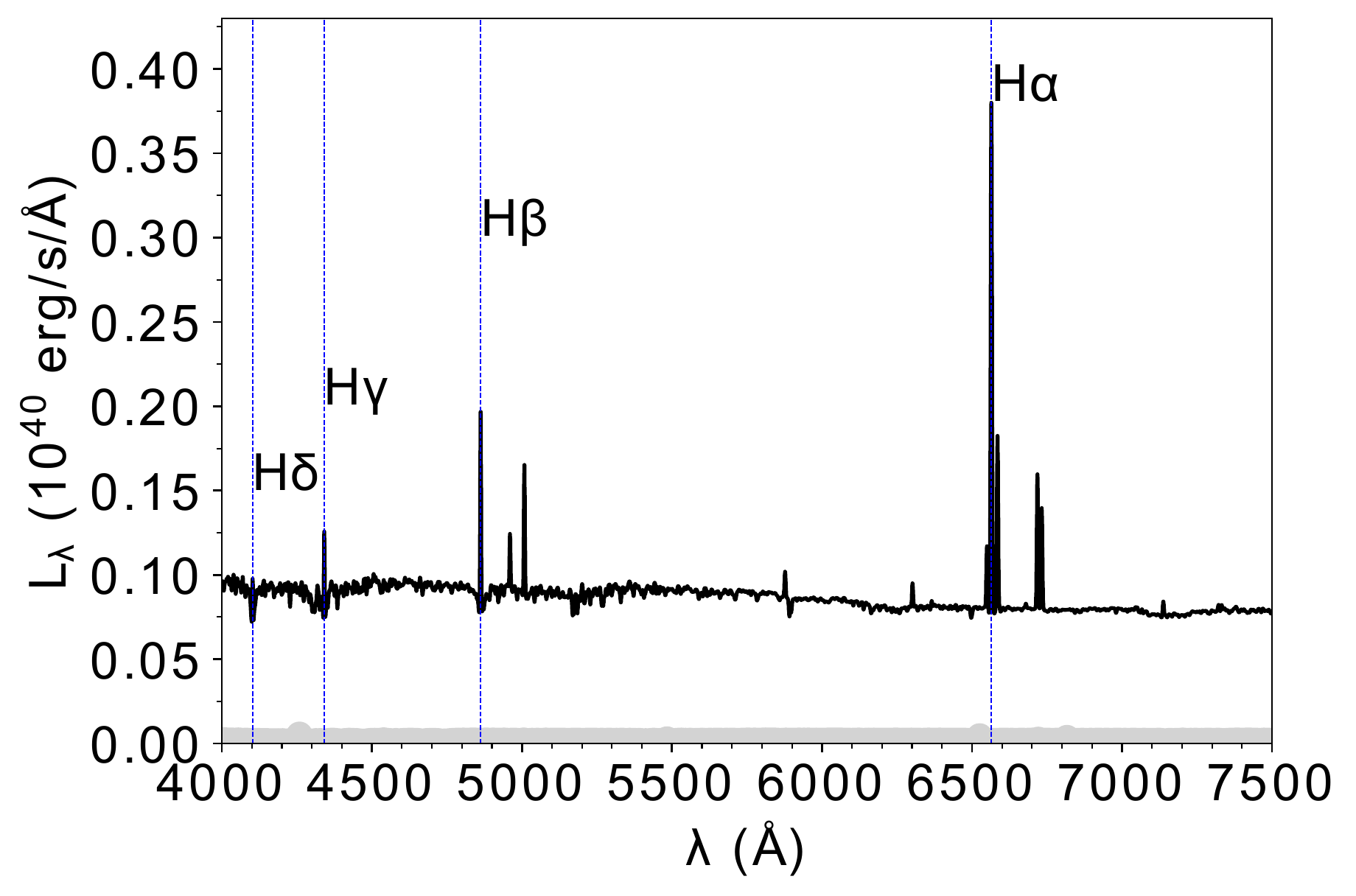}
    \caption{The composite spectrum constructed for all the  galaxies in sample shown in black. The grey region indicates the $\pm1\sigma$ uncertainty  in the spectrum. $\ha$, $\hb$, $\hg$, and $\hd$  emission lines are indicated by blue dotted lines.}
    \label{fig:comp}
\end{figure}

\section{\bf{composite spectrum}}
\label{sec:composite}
\subsection{Methodology of Constructing the Composite Spectrum}
\label{sec:compos}

We use composite spectra in order to measure the weaker Balmer lines including $\hg$ and $\hd$, which are typically not detected in the spectra of individual galaxies. The composite spectra are constructed by averaging, or stacking, the spectra of individual galaxies using the procedures given in \citetalias{Reddy_2020} and \textbf{specline}\footnote{\url{https://github.com/IreneShivaei/specline/}}~\citep{Shivaei18}. In brief, the science and error spectrum of each galaxy are shifted to the rest-frame based on the spectroscopic redshift, converted to luminosity density, and interpolated to a wavelength grid with spacing of $0.4$ \AA. The composite spectrum at each wavelength is calculated as an average of the luminosity densities of individual spectra that are weighted by their inverse variance. The error in the composite spectrum is derived using bootstrap resampling, where we randomly selected 2000 objects from the sample, perturbed their spectra according to the error spectra, and reconstructed the composite spectrum from these realizations. This process is repeated many times, and the resulting standard deviation in luminosity densities at each wavelength point gives the composite error spectrum. Figure~\ref{fig:comp} shows the composite spectrum and its error constructed for the $78,340$ objects in the sample.

\begin{figure}
\includegraphics[width=1\columnwidth]{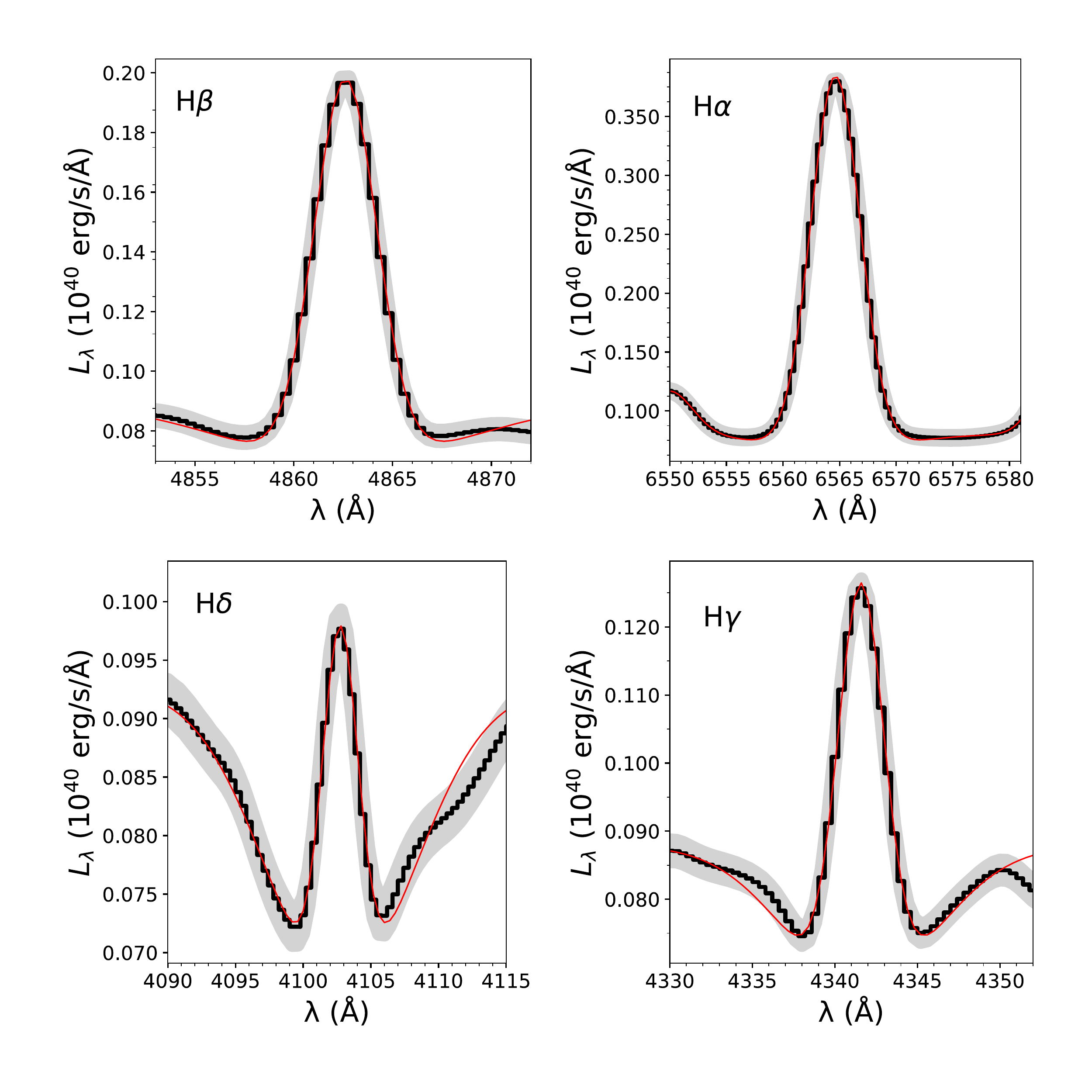}
\centering
\caption{$\ha,\hb,\hg,\hd$ emission lines observed in the composite spectrum of all galaxies in the sample, shown in black. The red lines show the best-fit Gaussian models that account for both emission and absorption for each line. The gray filled bands show the $1\sigma$ uncertainty of the spectrum.}
\label{fig:emissionlines}
\end{figure}

\begin{table}
\centering
\begin{threeparttable}[b]
\caption{Luminosity ($L$) measurements}
\label{tab:obsflux}
\def\arraystretch{1.3}%
\begin{tabular*}{\columnwidth}{@{\extracolsep{\fill}} lcc}
\hline
Line\tnote{a}& ${L}(10^{40}\mathrm{erg/s})$\tnote{b}&Fitting Window (\AA)\tnote{c}\\
\hline
\ha & $1.743\pm 0.0017$& $6442 - 6692$ \\
\hb & $0.5062\pm 0.0026$& $4813 - 4913$ \\
\hg & $0.2088\pm 0.0010$& $4265 - 4416$ \\
\hd & $0.1040\pm 0.0016$& $4015 - 4200$ \\
\hline
\end{tabular*}
\begin{tablenotes}
\item[a] Balmer Recombination Lines
\item [b] Luminosity and its error measured from the composite spectrum. Error in the line luminosity measured using the Monte Carlo method discussed in Section~\ref{sec:lineflux}.
\item [c] Wavelength range over which the lines are fit.
\end{tablenotes}
\end{threeparttable}
\end{table}

\subsection{Measurements of the Balmer emission lines from the composite spectrum}
\label{sec:lineflux}
$\ha,\hb,\hg$, and $\hd$ emission lines (Figure~\ref{fig:emissionlines}) are measured from the stacked spectrum. We chose not to include the $\he$ emission line ($\lambda = 3971.20$ \AA) in our analysis as it is blended with, and not well-resolved from, the $\rm [\ion{Ne}{III}]\lambda 3969$  line.

All lines have been measured by fitting two Gaussian functions, one to the absorption and one to the emission line except for the $\ha$ line. $\ha$ is fit simultaneously along with the $\rm [\ion{N}{II}]$ doublet and the underlying Balmer absorption. The velocity widths used to fit the $\hb$, $\hg$, and $\hd$ emission lines were constrained to be within the $20\%$ of the width obtained for $\ha$. The Balmer absorption measured from the composite spectrum is consistent with those inferred from the stellar population models (\citealt{Bruzual_Charlott2003}, $Z=0.020$ “solar”) that best fit the broadband photometry of galaxies contributing to the composite spectrum. The luminosity uncertainties are calculated by perturbing the stacked spectrum according to its error spectrum and remeasuring the line luminosities many times using the same method described in this section. The standard deviation of the values obtained in these iterations is adopted as the luminosity error. Table~\ref{tab:obsflux} reports the measured line luminosities from the composite spectrum for the entire sample.

\section{\bf Shape of the Nebular Attenuation Curve}
\label{sec:calculations}
\subsection{Definitions}
\label{sec:definition}
Here we discuss the methodology for determining the shape of the nebular attenuation curve. The intrinsic Balmer line ratios reported in Table~\ref{tab:ratio} are well determined and depend weakly on the local conditions such as electron density and temperature. The typical conditions assumed for the intrinsic $\ha/\hb$ ratio are  $n_{e}=100 \ \mathrm{cm^{-3}}$ and ${T_{e}=10000 \ \mathrm{K}}$ \citep{Osterbrock}. The relationship between the observed luminosity, $L(\lambda)$, and the intrinsic luminosity, $L_{0}(\lambda)$, can be expressed as follows:

\begin{eqnarray}
\label{eq:DefA}
L(\lambda)=L_{0}(\lambda) \times 10^{-0.4A(\lambda)},
\end{eqnarray}
where $A(\lambda)$ is the attenuation in magnitudes at wavelength $\lambda$. The total nebular dust attenuation curve is defined as $k(\lambda)$:

\begin{eqnarray}
k(\lambda) = \frac{A(\lambda)}{E(B-V)_{\rm neb}},
\label{eq:klam}
\end{eqnarray}
where $E(B-V)_{\rm neb}=A(B)-A(V)$ is defined as the color excess . The $B$ and $V$ bands are taken to be at $4400$ \AA \ and $5500$ \AA , respectively.

\begin{table}
\centering
\begin{threeparttable}[b]
\caption{Balmer Line Ratios}
\label{tab:ratio}
\def\arraystretch{1.3}%
\begin{tabular*}{\columnwidth}{@{\extracolsep{\fill}} lcc}
\hline
Line\tnote{a}& ${\lambda}$ (\AA)\tnote{b}&Line Ratios (\AA)\tnote{c}\\
\hline
$\ha$ &$6564.60$ & $2.860$ \\
$\hb$ &$4862.71$ & $1.000$ \\
$\hg$ &$4341.69$ & $0.468$ \\
$\hd$ &$4102.89$ & $0.259$ \\
\hline
\end{tabular*}
\begin{tablenotes}
\item [a] Balmer Recombination Lines.
\item [b] Rest-frame Vacuum wavelength.
\item [c] Intensity of line relative to $\hb$ for Case B recombination, $n_{\rm e}=10^2 \ \mathrm{cm^{-3}}$ and ${T_{\rm e}=10^4 \ \mathrm{K}}$ \citep{Osterbrock}.
\end{tablenotes}
\end{threeparttable}
\end{table}

\begin{figure*}
   \subfloat{%
      \includegraphics[width=1\columnwidth]{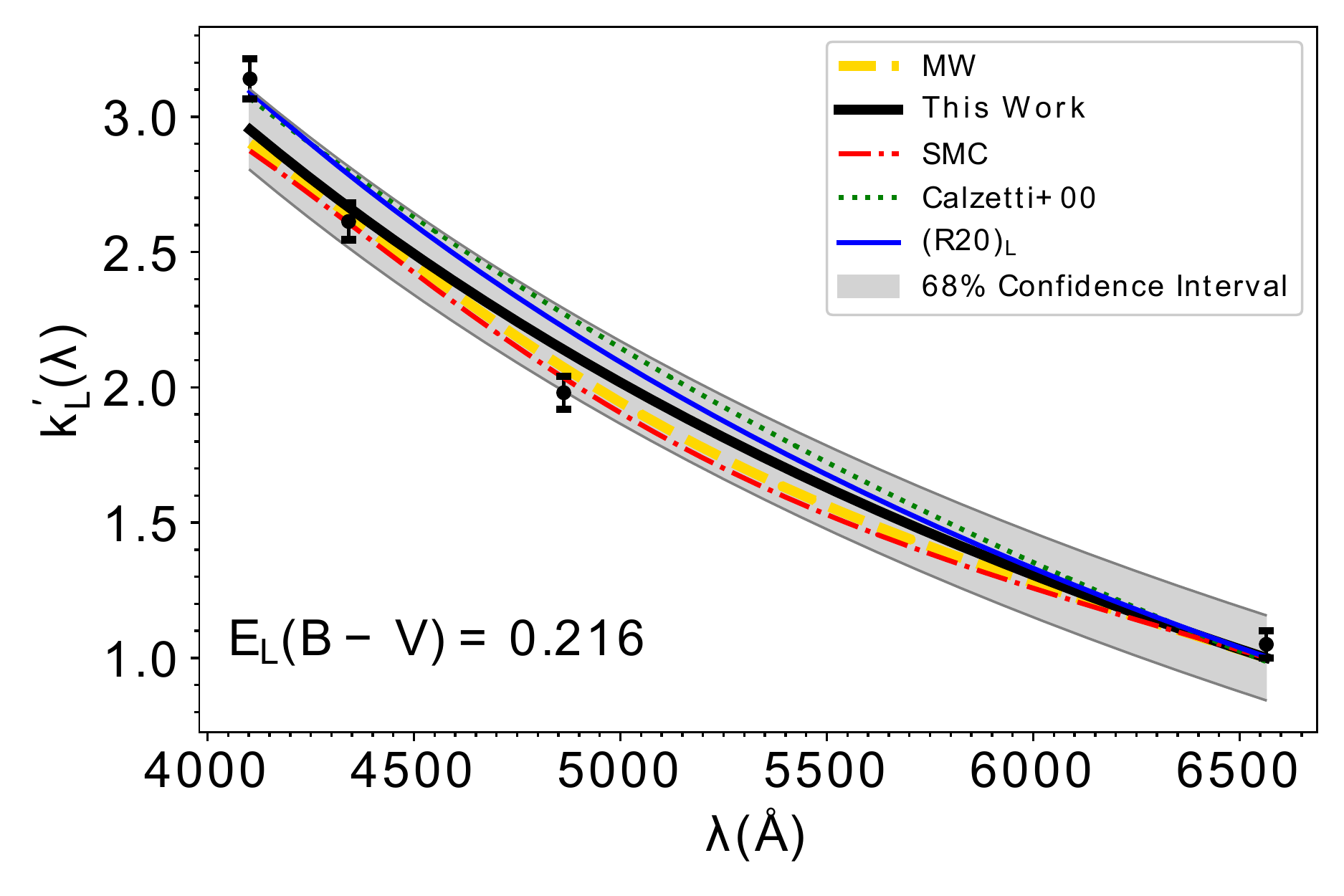}}
   \subfloat{%
      \includegraphics[width=1\columnwidth]{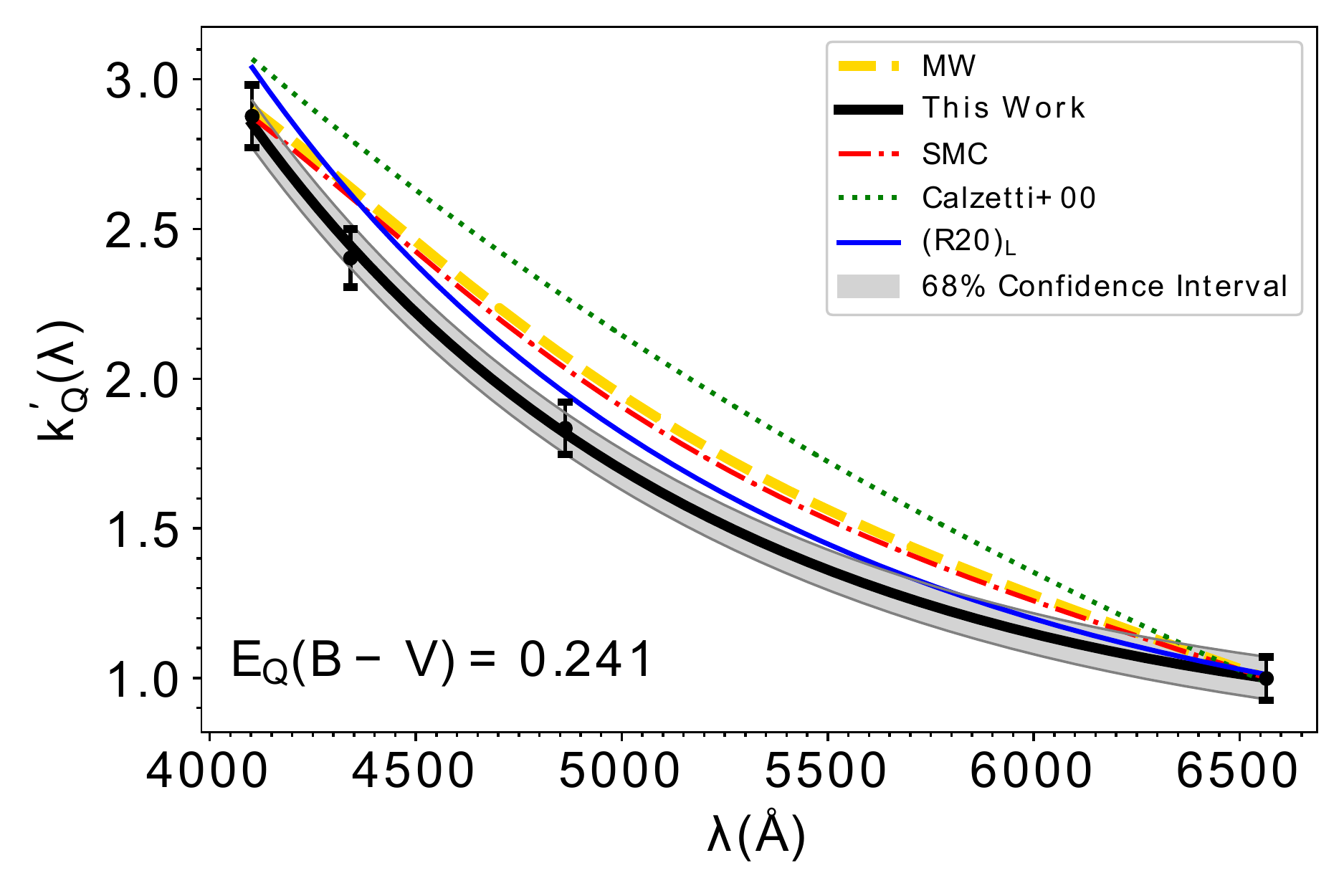}}
\caption{\label{fig:kav} Average nebular dust attenuation curve, $k'(\lambda)$,  versus $\lambda$, for the linear (left) and quadratic (right) polynomial forms. Attenuation curve measurements are shown by black circles along with their error bars. The best fit $68\%$ confidence intervals are shown by grey region and grey lines. For comparison, the MW extinction curve, SMC, \citet{Calzetti00} and the curves derived in \citetalias{Reddy_2020} are shown and  have been shifted, to have the same value at the wavelength of $\ha$ as the curves derived here. The subscripts used for \citetalias{Reddy_2020} refers to the curves based on fitting a linear or quadratic function. As it is indicated in the figure above, the reddening, $A'(4400$ \AA)$-A'(5500$ \AA), calculated by the linear form of the $A'(\lambda)$ is $\sim 10\%$ smaller than the one obtained by the quadratic form.}
\end{figure*}

\subsection{Methodology}
\label{sec:methodology}

We use the methodology introduced by \citetalias{Reddy_2020}  to calculate the shape of the nebular attenuation curve. In brief, \citetalias{Reddy_2020} expressed the attenuation in magnitudes relative to $\ha$ as follows:
\begin{eqnarray}
A'(\lambda) = & 2.5\left[\log_{\rm 10}\left(\frac{L(\ha)}{L(\lambda)}\right) - \log_{\rm 10}\left(\frac{L_{\rm 0}(\ha)}{L_{\rm 0}(\lambda)}\right)\right] + 1,
\label{eq:aprime}
\end{eqnarray}
where $L(\ha)/L(\lambda)$  is the observed ratio of the $\ha$ luminosity to that of a higher-order Balmer line ($\hb$, $\hg$, $\hd$), $L_{0}(\ha)/L_{0}(\lambda)$ denotes the intrinsic ratio, and $A'(\lambda)$ is equivalent to $A(\lambda)+[1-A(\ha)]$. The line luminosities measured from the composite spectrum are then used in conjunction with Equation~\ref{eq:aprime} to calculate the attenuation in magnitudes (relative to $\ha$) for each of the higher-order Balmer lines. We then fit linear and quadratic functions to $A'(\rm \lambda)$.

The shape of the attenuation curve, $k'(\lambda)$, can be related to $A'(\lambda)$ as follows:

\begin{eqnarray}
k'(\lambda)& \equiv & \frac{A'(\lambda)}{A'(4400 \ \angstrom)-A'(5500 \ \angstrom)}\nonumber \\
& = & k(\lambda) + \frac{[1-A(\ha)]}{E(B-V)_{\rm neb}}.
\label{eq:kplam}
\end{eqnarray}
Note that $k'(\lambda)$ and $k(\lambda)$ differ by an offset of $[1-A(\ha)]/\ebmvgas$ which is independent of $\lambda$. Therefore, $k'(\lambda)$ and $k(\lambda)$ are equivalent except for a normalization factor. In order to calculate $k'(\lambda)$, $A'(4400$ \AA)$-A'(5500$ \AA) is determined using linear-in-$1/\lambda$ (${A'(\lambda)=a_{0}+a_{1}/\lambda}$)  and quadratic-in-$1/\lambda$ ($A'(\lambda)=a_{0}+a_{1}/\lambda +a_{1}/\lambda^{2}$) fits to $A'(\lambda)$, and then $k'(\lambda)$ is computed using Equation~\ref{eq:kplam}. Next, we use the linear and quadratic polynomial forms discussed above to fit $k'(\lambda)$ vs. $\lambda$. More complicated functional forms are not considered due to the limited number of data points available to derive the attenuation curve. The uncertainty in a given $k'(\lambda)$ point is propagated throughout these calculations. The line ratios measurements are perturbed according to their errors, then $A'(\lambda)$, $\ebmvgas$ and $k'(\lambda)$ are recalculated many times to then determine the propagated measurement uncertainty in a given $k'(\lambda)$ point. The functional forms of the attenuation curves are:
\begin{eqnarray}
k^\prime_{\rm L}(\lambda) & = & -2.253 + \frac{2.135}{\lambda},
\label{eq:kcconstlin}
\end{eqnarray}

\begin{eqnarray}
k^\prime_{\rm Q}(\lambda) & = & 2.705 - \frac{3.083}{\lambda} + \frac{1.290}{\lambda^2},
\label{eq:kcconst}
\end{eqnarray}
where $\lambda$ is in units of $\mu$m,
in the range $0.41\le \lambda\le 0.66$\,$\mu$m. Note that $k'_{\rm L}(\lambda)$ and $k'_{\rm Q}(\lambda)$ denote the curves based on fitting a linear-in-$1/\lambda$ and quadratic-in-$1/\lambda$ function, respectively, and the curves are all normalized such that their values at the wavelength of $\ha$ is equal to one to aid in comparing them with other curves in the literature. We consider both the linear and quadratic functions to demonstrate the associated systematic uncertainty in the resulting nebular attenuation curve.

The nebular attenuation curve derived here is shown in Figure~\ref{fig:kav}. The Galactic extinction curve \citep{Cardelli89}, \citet{Calzetti00} curve, SMC, and the nebular curves derived for redshift $z\sim2$ galaxies in \citetalias{Reddy_2020} are also shown in Figure~\ref{fig:kav}. The MW curve is typically used for the extinction correction of nebular lines, while \citet{Calzetti00} and SMC are often used for the reddening of the stellar continuum in high-redshift galaxies. Figure~\ref{fig:kav} shows that the average nebular dust attenuation curve derived for low-redshift star-forming galaxies is similar to the nebular curves presented in \citetalias{Reddy_2020}  within $\mathrm{ 1\sigma}$, the MW and SMC curves within $\mathrm{ 2\sigma}$ confidence. These results imply that the combined effects of dust properties and geometry yield a shape of the curve that is similar to other common extinction and attenuation curves at rest-frame optical wavelengths.  We do not have sufficient information to disentangle changes in dust properties and geometry, and radiation transfer models indicate that curves of similar shape can be produced by dust distributions with substantially different properties (e.g., \citealt{Witt,Seon_2016}). Note that there are small differences in $k'_{\rm L}(\lambda)$ and $k'_{\rm Q}(\lambda)$ because the values of  $A'(4400$ \AA) and  $A'(5500$ \AA) depend on the functional form (i.e., linear or quadratic) used to determine these values.

To obtain the normalized total nebular dust attenuation curve $k(\rm \lambda)$ (Equation~\ref{eq:klam}) from $k'(\rm \lambda)$ (Equation~\ref{eq:kplam}), $k'_{\rm L}(\rm \lambda)$ is extrapolated to $\lambda = 2.8$\,$\mu$m, corresponding to the wavelength at which other common curves (e.g., MW, SMC, and LMC) approach zero \citep{Gordon03,Reddy_2015}. Similarly, $k_{\rm Q}(\rm \lambda)$ is assumed to have the same functional behavior as $k_{\rm L}(\rm \lambda)$ at long wavelength. Therefore, $k_{\rm Q}(\rm \lambda)$ is normalized such that it is equal to $k_{\rm L}(\rm \lambda)$ at $\lambda = 0.66$\,$\mu$m in order to obtain a continuous function. The final form of $k(\rm \lambda)$ is:

\begin{eqnarray}
k_{\rm L}(\lambda) & = & -0.762 + \frac{2.135}{\lambda}, \nonumber \\
& & 0.41 \le \lambda \le 0.66\, \mu{\rm m}.
\label{eq:kclin}
\end{eqnarray}

\begin{eqnarray}
k_{\rm Q}(\lambda) & = & 4.182 - \frac{3.083}{\lambda} + \frac{1.290}{\lambda^2}, \nonumber \\
& & 0.41 \le \lambda \le 0.66\, \mu{\rm m}; \nonumber \\
& = & -0.762 + \frac{2.135}{\lambda}, \nonumber \\
& & \lambda > 0.66\, \mu{\rm m}.
\label{eq:kcquad}
\end{eqnarray}

The total to selective absorption ratio is $R_{\rm V}=3.12$ and $2.84$ for the linear and quadratic forms, respectively. There are two sources of systematic uncertainty in $R_{\rm V}$. One is associated with the functional form used to fit the nebular attenuation curve. This error can be estimated by the difference in the values of $R_{\rm V}$ obtained for $k_{\rm Q}$ and $k_{\rm L}$ as $\Delta R\simeq0.28$. The other systematic error in $R_{\rm V}$ originates from utilizing different normalization methods, for example, using another value of the wavelength (rather than $2.8$\,$\mu$m) to set the nebular attenuation curve to zero \citep{Reddy_2020}. This error is $\delta R\simeq0.05$ if we set the zero-point to $3\,\mu$m instead. Overall, our result here is consistent within the uncertainty with the $R_{\rm V}=3.1$ reported by \citet{Cardelli89} for the average MW curve, $R_{\rm V}=2.9$ and $2.74$ reported by \citet{Pei} and \citet{Gordon03} for the SMC curve.  Our results are also consistent with the $R_{\rm V}$ values reported for the two  linear-in-$1/\lambda$ and quadratic-in-$1/\lambda$ attenuation curves derived in \citetalias{Reddy_2020} ($R_{\rm V}=3.34$ and $3.09$, respectively).

\begin{table*}

\setlength{\tabcolsep}{12pt}
\begin{threeparttable}[b]
\caption{Properties of subsamples, Reddenings, $R_{\rm V}$}
\label{tab:subsamples}
\def\arraystretch{1.3}%
\begin{tabular*}{\textwidth}{@{\extracolsep{\fill}} lc c c c c c}
\hline
Property \tnote{a} & Bin Range \tnote{b} & Median \tnote{c}& $E_{\rm L}(B-V)$\tnote{d}&$E_{\rm Q}(B-V)$\tnote{e}&$(R_{\rm V})_{\rm L}$\tnote{f} &$(R_{\rm V})_{\rm Q}$\tnote{g}\\
\hline
\vspace{2pt}
$\log (M_{\ast}/M_\odot)$& 6.68 , 8.94 &$ 8.71$ &$ 0.059 \pm  0.004$&$0.064 \pm 0.006$&$3.189 \pm 0.263$&$2.931 \pm 0.263$ \\
\vspace{2pt}
 & 8.94 , 9.23 & $   9.10$ & $0.179 \pm 0.003$ &$0.201 \pm 0.005$&$3.117 \pm 0.306$&$2.815 \pm 0.306$ \\
\vspace{2pt}
 & 9.23 , 9.48 & $  9.36$ &$0.292 \pm 0.003$ &$0.323 \pm 0.005$&$3.106 \pm 0.274$&$2.837 \pm 0.274$ \\
\vspace{2pt}
 & 9.48 , 9.75 & $   9.61$ &$0.381 \pm 0.004$&$0.432 \pm 0.007 $& $3.040 \pm 0.305$& $2.739 \pm 0.305$ \\
\vspace{2pt}
& 9.75 , 11.46 & $   9.92$& $0.608 \pm 0.007$&$ 0.672 \pm 0.012$&$3.036 \pm 0.269$&$2.772 \pm 0.269$ \\
\hline
\vspace{2pt}
$\log (sSFR/ \rm yr^{-1})$ & $-11.79$ , $-10.02$&$ -10.15$ &$0.452 \pm 0.020$ &$0.525 \pm 0.034$& $2.948 \pm 0.374$&$2.613 \pm 0.374$ \\
\vspace{2pt}
& $-10.02$ , $-9.83$&$  -9.92$&$0.377 \pm 0.011$ &$0.428 \pm 0.019$&$3.041 \pm 0.305$&$2.740 \pm 0.305$ \\
\vspace{2pt}
& $-9.83$ , $-9.67$ & $ -9.75$ &$ 0.338 \pm 0.008$ &$0.380 \pm 0.014$&$3.062 \pm 0.285$&$2.781 \pm 0.285$\\
\vspace{2pt}
& $-9.67$ , $-9.46$ & $  -9.58$&$0.349 \pm 0.007$ &$0.341 \pm 0.012$&$3.132 \pm 0.327$&$2.809 \pm 0.327$ \\
\vspace{2pt}
& $-9.46$ , $-7.05$ & $  -9.29$ &$0.179 \pm 0.006$ &$0.191 \pm 0.010$&$3.145 \pm 0.302$&$2.847 \pm 0.302$\\
\hline
\vspace{2pt}
$12+\log(\rm{O}/\rm{H})$ &$7.85$ , $8.73$ & $ 8.63$ & $ 0.059 \pm 0.007$ &$0.068 \pm 0.011 $&$3.162 \pm 0.372$&$2.793 \pm 0.372$ \\
\vspace{2pt}
   & $8.73$ , $8.87$ & $  8.82$ &$0.180 \pm 0.003$ &$0.207 \pm 0.004$&$3.089 \pm 0.349$ &$2.744 \pm 0.349$ \\
\vspace{2pt}
   &  $8.87$ , $8.98$ & $  8.94$&$0.288 \pm 0.008$ &$0.331 \pm 0.014$&$3.046 \pm 0.332$ &$2.718 \pm 0.332$ \\
\vspace{2pt}
   & $8.98$ , $9.06$ & $  9.01$& $ 0.413\pm 0.010$ &$ 0.468 \pm 0.017$&$3.038 \pm 0.306$&$2.736 \pm 0.306$ \\
\vspace{2pt}
   & $9.06$ , $9.40$ & $  9.11$ &$0.587 \pm 0.013$ &$0.668 \pm 0.023$&$2.984 \pm 0.331$ &$2.657 \pm 0.331$ \\
\hline
\end{tabular*}
\begin{tablenotes}
\item [a] First, second, and third five rows indicate bins in stellar mass, sSFR, and metallicity, respectively. Sample size is 15,668 galaxies in each bin.
\item [b] The range of the associated physical property in each bin.
\item [c] Median value of the associated physical property in each bin.
\item [d] Reddening computed from the linear form of $A'(\lambda)$ (Equation~\ref{eq:aprime}).
\item [e] Reddening computed from the quadratic form of $A'(\lambda)$ (Equation~\ref{eq:aprime}).
\item[f] Total to selective absorption ratio calculated using the linear form of the total nebular dust attenuation curve.
\item[g] Total to selective absorption ratio calculated using the quadratic form of the total nebular dust attenuation curve.
\end{tablenotes}
\end{threeparttable}
\end{table*}

\begin{figure}
    \includegraphics[width=1\columnwidth]{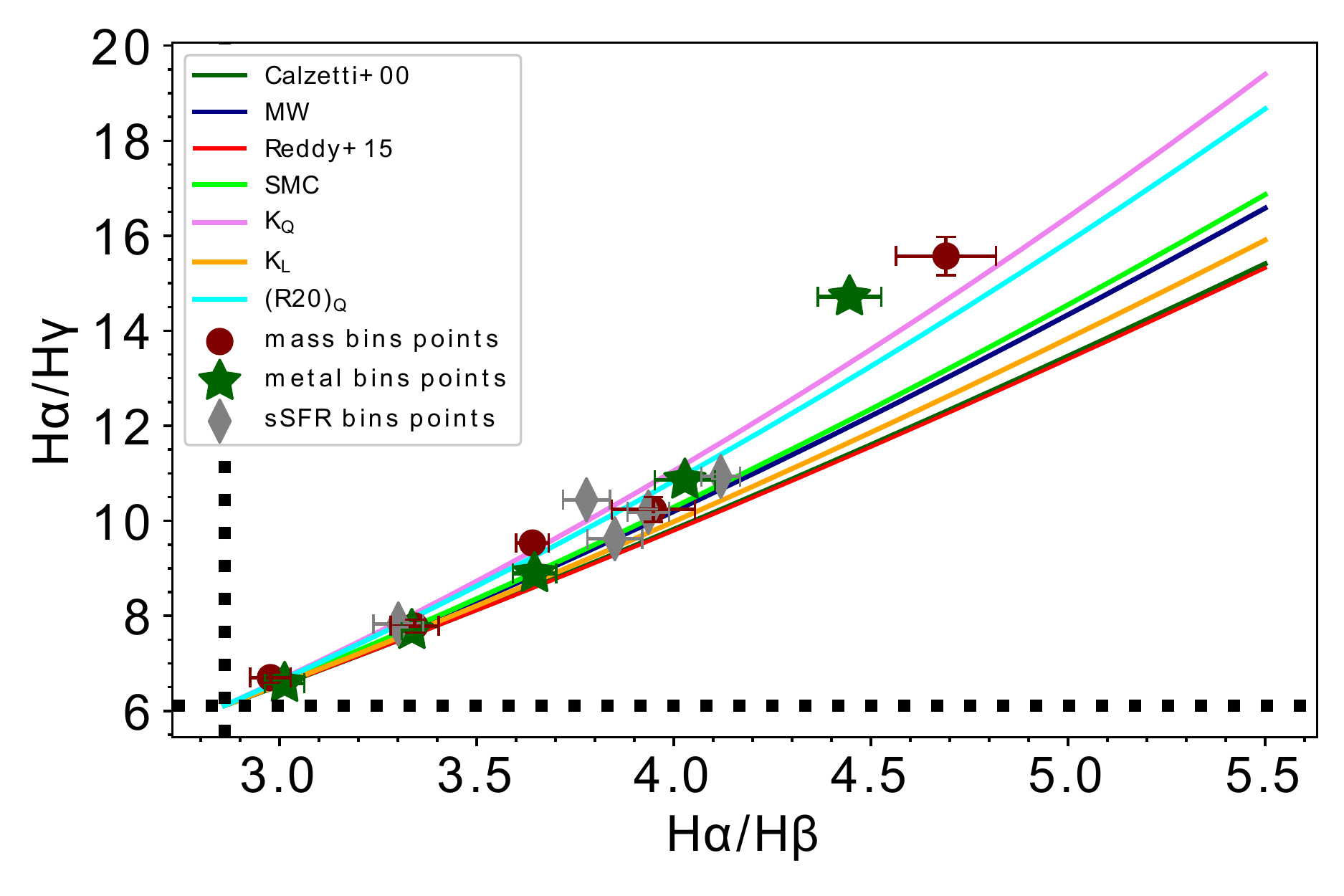}
    \caption{Ratios of $\ha/\hg$ vs $\ha/\hb$ for stellar mass, metallicity, and sSFR bins. The error bars are also indicated for each point. The dotted black lines indicate the intrinsic line ratios. The relationship between these line ratios for different extinction/attenuation curves are indicated by the curves.}  
    \label{fig:hahg}
\end{figure}

\begin{figure*}

   \subfloat{%
      \includegraphics[width=1\columnwidth]{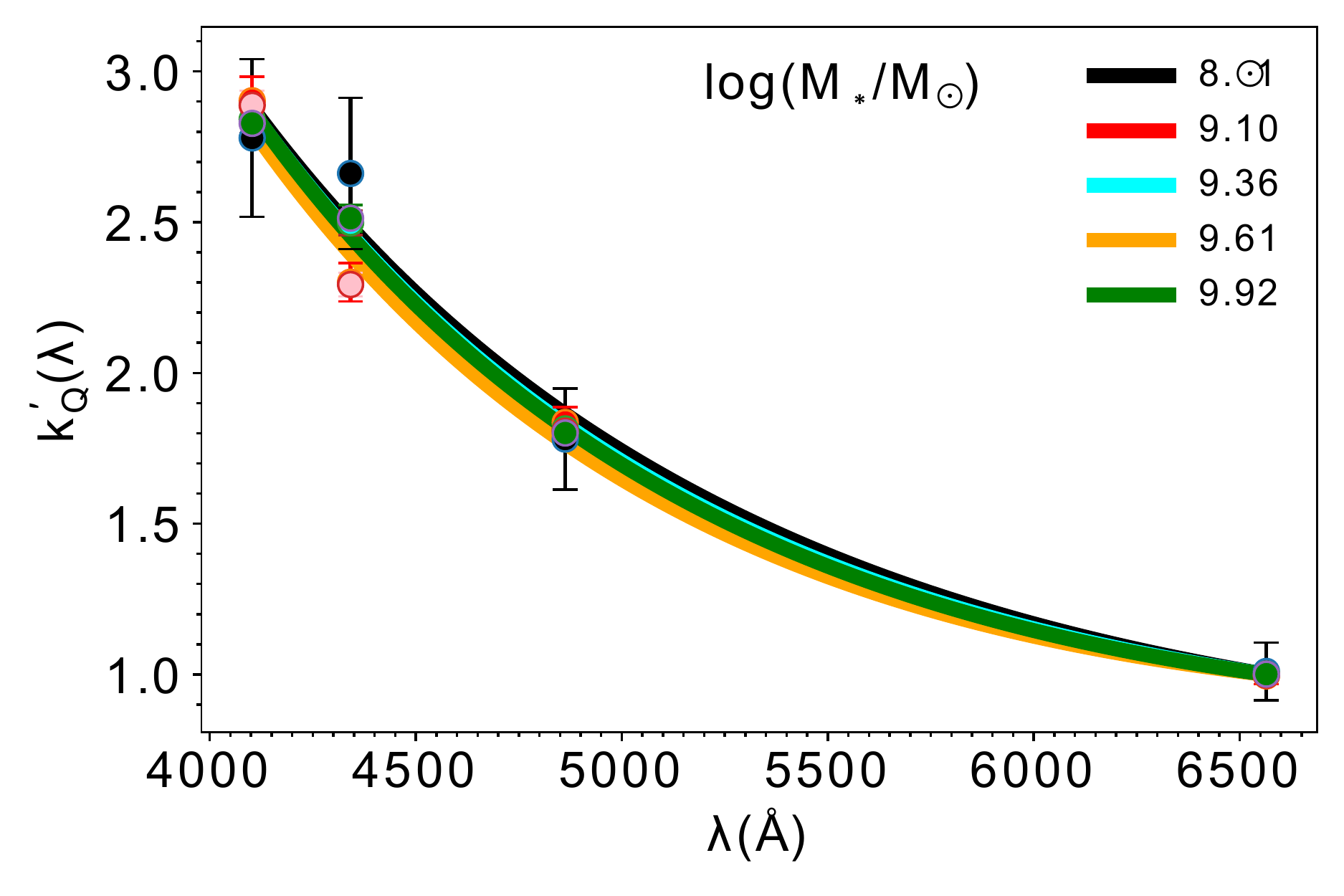}}

   \subfloat{%
      \includegraphics[width=1\columnwidth]{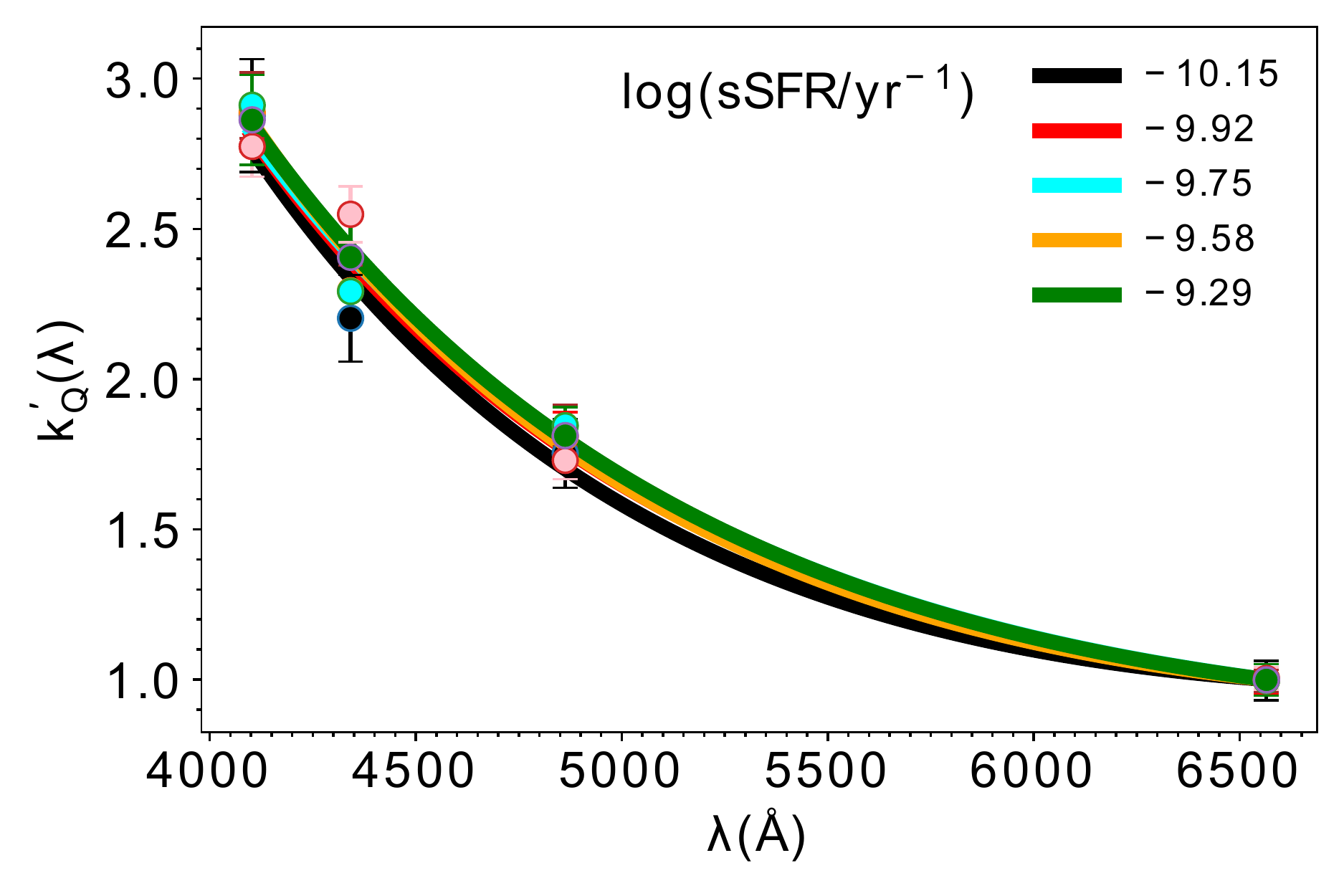}}

   \subfloat{%
      \includegraphics[width=1\columnwidth]{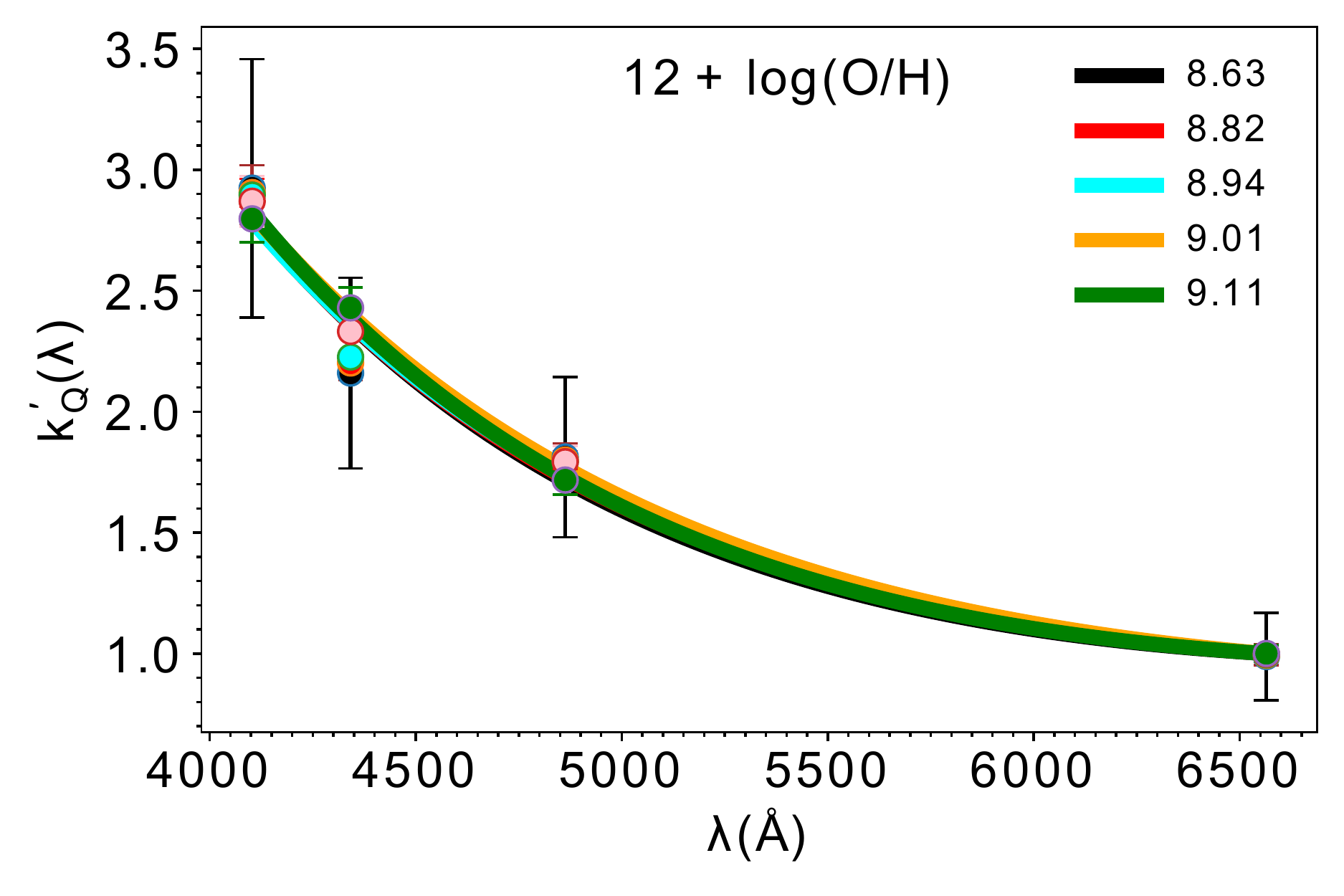}} \\
\caption{\label{fig:var} Top, middle, and bottom panels indicate the curves derived for each bin of stellar mass, sSFR, and metallicity, respectively. The curves have been shifted so that their values at the wavelength of $\ha$ are set equal to one. The attenuation curve points are shown by colored circles along with their error bars. The median values of the physical property in each of the bins are shown in the top right corner of each panel.}
\end{figure*}

\section{\bf {Nebular Attenuation Curve vs. Galaxy Properties}}
\label{sec:subsample}
To examine whether the curve varies with galaxy properties, we subdivide our sample into five bins each of stellar mass, sSFR, and metallicity, with each containing $15,668$ galaxies. Composite spectra are constructed for each of the subsamples following the method outlined in Section~\ref{sec:compos}, and we use the same methodology outlined in \ref{sec:methodology} to derive the nebular attenuation curve. Table~\ref{tab:subsamples} reports the physical properties of galaxies in each of the subsamples, and the properties of the derived nebular attenuation curves for each of the bins.

Figure~\ref{fig:hahg} shows $\ha/\hg$ versus $\ha/\hb$ measured from the stellar mass, metallicity and sSFR bins. All the bins are consistent with $k_{\rm Q}$ curve within their $1 \sigma$ uncertainties, except for the one bin with the highest metallicity that covers the $k_{\rm Q}$ curve within its $2 \sigma$ uncertainty, which is reasonable enough that we cannot rule out $k_{\rm Q}$ curve for this bin. In comparison to $k_{\rm L}$, $k_{\rm Q}$ is found to best match the line ratio measurements. This is not particularly surprising given the additional free parameter of the quadratic fit versus the linear fit. The higher-order polynomial functional form reflects the wavelength behavior exhibited by the other common extinction and attenuation curves (e.g., \citealt{Cardelli89,Calzetti00,Gordon03}),\footnote{For the longer-wavelength ($\lambda > 7000$ \AA ) their shape is typically characterized by an inverse power-law in $\lambda$.} which is another reason why $k_{\rm Q}$ is preferred in this analysis. 

The nebular attenuation curves ($k'_{\rm Q}$) derived for the bins of stellar mass, metallicity, and sSFR are shown in Figure~\ref{fig:var}. The curves in each panel of Figure~\ref{fig:var} are consistent with each others within their $1\sigma $ confidence interval, suggesting that the shape of the nebular attenuation curve shows little to no variations when binned by stellar mass, sSFR, and metallicity. The ratio of the total to selective absorption at V-band ($R_{\rm V}$) are computed for each of the curve fits (Table~\ref{tab:subsamples}). The $R_{\rm V}$ values for the curves in each associated physical property are consistent with each others within their $1\sigma$ systematic uncertainties. This implies that the normalization of the nebular attenuation curve does not vary with the aforementioned properties.

\section{\bf Discussion}
\label{sec:discuss}

Our results indicate that there is no significant variation in the shape of the nebular attenuation curve in the range of $0.41\,\mu$m to $0.66\,\mu$m with stellar mass, metallicity, and sSFR for low-redshift star-forming galaxies. This lack of variation may be related to the fact that the nebular attenuation curve is only probing those sightlines towards massive stars, where in a simplified scenario, the dust configuration can be approximated as a foreground screen and the dust size distribution is dictated by the radiation field of the youngest stellar populations.  Because of the latter, one might not expect much variation in the shape of the curve as a function of globally-derived properties that are not solely sensitive to the youngest stellar populations. Although, there still exists the possibility that identical curves in the optical regions can be attributed to dust distribution with different properties.

The overall combination of the dust geometry and composition dictates the shape of the nebular attenuation curve. In addition to that, the connection between the extinction and attenuation curves can be complicated. Therefore, it is difficult to identify any particular similarities in the dust properties of the various bins solely based on the fact that the attenuation curves have identical shapes.


\section{\bf Summary}
\label{sec:con}
We use $78,340$ spectra of local star-forming galaxies with the redshift range of $z\simeq 0.04-0.1$ to investigate whether the shape of the nebular attenuation curve varies with the inferred physical properties of the sample. We use the first four detected Balmer lines ($\ha, \hb, \hg, \hd$) from the stacked spectrum of all the galaxies in the sample to derive an average nebular attenuation curve using linear and quadratic polynomial functional forms in terms of $1/\lambda$.

The curves derived in this work are consistent with the nebular attenuation curves presented in \citetalias{Reddy_2020} for high-redshift galaxies within $1\sigma$ and the MW and SMC curves within  $2\sigma$ confidence interval. The $R_{\rm V}$ values obtained for the curves derived in this work are consistent with the ones computed for the Galactic extinction and SMC curves, and the curves presented in \citetalias{Reddy_2020}, showing that the curves are also similar to that of the MW, SMC, and nebular curves derived in \citetalias{Reddy_2020} in terms of the normalization. 

We calculate the nebular attenuation curve for galaxies in bins of stellar mass, metallicity, and sSFR, and compare their shapes. The curves derived in these various bins are identical to each other within the uncertainties.

The analysis outlined here may be extended to also examine the nebular curve in galaxies hosting AGN, and to determine if the presence of the hard radiation field of AGN may influence dust grain size distributions and/or geometry.

\section{\bf Acknowledgement}
We acknowledge that we have used the publicly published data from the SDSS survey. Funding for SDSS-III has been provided by the Alfred P. Sloan Foundation, the Participating Institutions, the National Science Foundation, and the U.S. Department of Energy Office of Science. The SDSS-III web site is \url{http://www.sdss3.org/.} SDSS-III is managed by the Astrophysical Research Consortium for the Participating Institutions of the SDSS-III Collaboration, including the University of Arizona, the Brazilian Participation Group, Brookhaven National Laboratory, Carnegie Mellon University, University of Florida, the French Participation Group, the German Participation Group, Harvard University, the Instituto de Astrofisica de Canarias, the Michigan State/Notre Dame/JINA Participation Group, Johns Hopkins University, Lawrence Berkeley National Laboratory, Max Planck Institute for Astrophysics, Max Planck Institute for Extraterrestrial Physics, New Mexico State University, New York University, Ohio State University, Pennsylvania State University, University of Portsmouth, Princeton University, the Spanish Participation Group, University of Tokyo, University of Utah, Vanderbilt University, University of Virginia, University of Washington, and Yale University.

We also acknowledge that Ali Ahmad Khostovan's research is supported by an appointment to the NASA Postdoctoral Program at the Goddard Space Flight Center, administered by the Universities Space Research Association (USRA) through a contract with NASA.

\section*{Data Availability}
The spectra of galaxies analyzed in this work are publicly available from the SDSS survey. We use the SDSS Data Release $8$ \citep{Aihara_2011} for the purpose of this work. We get the spectra of galaxies in our sample by submitting a query in~\url{https://skyserver.sdss.org/casjobs/}. The SDSS catalogs ($Galspec$) containing the basic information and line measurements for each spectrum are provided by MPA/JHU group \citep{Kauffmann_2003,Brinchmann_2004,Tremonti2004} and are available from: ~\url{http://www.sdss3.org/dr8/spectro/galspec.php}.

\bibliographystyle{mnras}
\bibliography{ms}

\begin{thebibliography}{}
\makeatletter
\relax
\def\mn@urlcharsother{\let\do\@makeother \do\$\do\&\do\#\do\^\do\_\do\%\do\~}
\def\mn@doi{\begingroup\mn@urlcharsother \@ifnextchar [ {\mn@doi@}
  {\mn@doi@[]}}
\def\mn@doi@[#1]#2{\def\@tempa{#1}\ifx\@tempa\@empty \href
  {http://dx.doi.org/#2} {doi:#2}\else \href {http://dx.doi.org/#2} {#1}\fi
  \endgroup}
\def\mn@eprint#1#2{\mn@eprint@#1:#2::\@nil}
\def\mn@eprint@arXiv#1{\href {http://arxiv.org/abs/#1} {{\tt arXiv:#1}}}
\def\mn@eprint@dblp#1{\href {http://dblp.uni-trier.de/rec/bibtex/#1.xml}
  {dblp:#1}}
\def\mn@eprint@#1:#2:#3:#4\@nil{\def\@tempa {#1}\def\@tempb {#2}\def\@tempc
  {#3}\ifx \@tempc \@empty \let \@tempc \@tempb \let \@tempb \@tempa \fi \ifx
  \@tempb \@empty \def\@tempb {arXiv}\fi \@ifundefined
  {mn@eprint@\@tempb}{\@tempb:\@tempc}{\expandafter \expandafter \csname
  mn@eprint@\@tempb\endcsname \expandafter{\@tempc}}}

\bibitem[\protect\citeauthoryear{{Aihara} et~al.,}{{Aihara}
  et~al.}{2011}]{Aihara_2011}
{Aihara} H.,  et~al., 2011, \mn@doi [\apjs] {10.1088/0067-0049/193/2/29}, \href
  {https://ui.adsabs.harvard.edu/abs/2011ApJS..193...29A} {193, 29}

\bibitem[\protect\citeauthoryear{{Battisti}, {Calzetti}  \& {Chary}}{{Battisti}
  et~al.}{2016}]{Battisti}
{Battisti} A.~J.,  {Calzetti} D.,   {Chary} R.~R.,  2016, \mn@doi [\apj]
  {10.3847/0004-637X/818/1/13}, \href
  {https://ui.adsabs.harvard.edu/abs/2016ApJ...818...13B} {818, 13}

\bibitem[\protect\citeauthoryear{{Battisti}, {Calzetti}  \& {Chary}}{{Battisti}
  et~al.}{2017}]{Battisti17}
{Battisti} A.~J.,  {Calzetti} D.,   {Chary} R.~R.,  2017, \mn@doi [\apj]
  {10.3847/1538-4357/aa6fb2}, \href
  {https://ui.adsabs.harvard.edu/abs/2017ApJ...840..109B} {840, 109}

\bibitem[\protect\citeauthoryear{{Bianchi}, {Clayton}, {Bohlin}, {Hutchings}
  \& {Massey}}{{Bianchi} et~al.}{1996}]{Bianch}
{Bianchi} L.,  {Clayton} G.~C.,  {Bohlin} R.~C.,  {Hutchings} J.~B.,   {Massey}
  P.,  1996, \mn@doi [\apj] {10.1086/177963}, \href
  {https://ui.adsabs.harvard.edu/abs/1996ApJ...471..203B} {471, 203}

\bibitem[\protect\citeauthoryear{{Brinchmann}, {Charlot}, {White}, {Tremonti},
  {Kauffmann}, {Heckman}  \& {Brinkmann}}{{Brinchmann}
  et~al.}{2004}]{Brinchmann_2004}
{Brinchmann} J.,  {Charlot} S.,  {White} S.~D.~M.,  {Tremonti} C.,  {Kauffmann}
  G.,  {Heckman} T.,   {Brinkmann} J.,  2004, \mn@doi [\mnras]
  {10.1111/j.1365-2966.2004.07881.x}, \href
  {https://ui.adsabs.harvard.edu/abs/2004MNRAS.351.1151B} {351, 1151}

\bibitem[\protect\citeauthoryear{{Bruzual} \& {Charlot}}{{Bruzual} \&
  {Charlot}}{2003}]{Bruzual_Charlott2003}
{Bruzual} G.,  {Charlot} S.,  2003, \mn@doi [\mnras]
  {10.1046/j.1365-8711.2003.06897.x}, \href
  {https://ui.adsabs.harvard.edu/abs/2003MNRAS.344.1000B} {344, 1000}

\bibitem[\protect\citeauthoryear{{Buat} et~al.,}{{Buat}
  et~al.}{2011}]{Buat2011}
{Buat} V.,  et~al., 2011, \mn@doi [\aap] {10.1051/0004-6361/201117264}, \href
  {https://ui.adsabs.harvard.edu/abs/2011A&A...533A..93B} {533, A93}

\bibitem[\protect\citeauthoryear{{Buat} et~al.,}{{Buat}
  et~al.}{2012}]{Buat2012}
{Buat} V.,  et~al., 2012, \mn@doi [\aap] {10.1051/0004-6361/201219405}, \href
  {https://ui.adsabs.harvard.edu/abs/2012A&A...545A.141B} {545, A141}

\bibitem[\protect\citeauthoryear{{Buat}, {Boquien}, {Ma{\l}ek}, {Corre},
  {Salas}, {Roehlly}, {Shirley}  \& {Efstathiou}}{{Buat} et~al.}{2018}]{Buat}
{Buat} V.,  {Boquien} M.,  {Ma{\l}ek} K.,  {Corre} D.,  {Salas} H.,  {Roehlly}
  Y.,  {Shirley} R.,   {Efstathiou} A.,  2018, \mn@doi [\aap]
  {10.1051/0004-6361/201833841}, \href
  {https://ui.adsabs.harvard.edu/abs/2018A&A...619A.135B} {619, A135}

\bibitem[\protect\citeauthoryear{{Calzetti}}{{Calzetti}}{1997}]{Calzetti97}
{Calzetti} D.,  1997, \mn@doi [\aj] {10.1086/118242}, \href
  {https://ui.adsabs.harvard.edu/abs/1997AJ....113..162C} {113, 162}

\bibitem[\protect\citeauthoryear{{Calzetti}}{{Calzetti}}{2001}]{Calzetti2001}
{Calzetti} D.,  2001, \mn@doi [\pasp] {10.1086/324269}, \href
  {https://ui.adsabs.harvard.edu/abs/2001PASP..113.1449C} {113, 1449}

\bibitem[\protect\citeauthoryear{{Calzetti}, {Kinney}  \&
  {Storchi-Bergmann}}{{Calzetti} et~al.}{1994}]{Calzetti94}
{Calzetti} D.,  {Kinney} A.~L.,   {Storchi-Bergmann} T.,  1994, \mn@doi [\apj]
  {10.1086/174346}, \href
  {https://ui.adsabs.harvard.edu/abs/1994ApJ...429..582C} {429, 582}

\bibitem[\protect\citeauthoryear{{Calzetti}, {Armus}, {Bohlin}, {Kinney},
  {Koornneef}  \& {Storchi-Bergmann}}{{Calzetti} et~al.}{2000}]{Calzetti00}
{Calzetti} D.,  {Armus} L.,  {Bohlin} R.~C.,  {Kinney} A.~L.,  {Koornneef} J.,
   {Storchi-Bergmann} T.,  2000, \mn@doi [\apj] {10.1086/308692}, \href
  {https://ui.adsabs.harvard.edu/abs/2000ApJ...533..682C} {533, 682}

\bibitem[\protect\citeauthoryear{{Cardelli}, {Clayton}  \& {Mathis}}{{Cardelli}
  et~al.}{1989}]{Cardelli89}
{Cardelli} J.~A.,  {Clayton} G.~C.,   {Mathis} J.~S.,  1989, \mn@doi [\apj]
  {10.1086/167900}, \href
  {https://ui.adsabs.harvard.edu/abs/1989ApJ...345..245C} {345, 245}

\bibitem[\protect\citeauthoryear{{Charlot} \& {Fall}}{{Charlot} \&
  {Fall}}{2000}]{Charlot}
{Charlot} S.,  {Fall} S.~M.,  2000, \mn@doi [\apj] {10.1086/309250}, \href
  {https://ui.adsabs.harvard.edu/abs/2000ApJ...539..718C} {539, 718}

\bibitem[\protect\citeauthoryear{{Chevallard}, {Charlot}, {Wandelt}  \&
  {Wild}}{{Chevallard} et~al.}{2013}]{Cheva}
{Chevallard} J.,  {Charlot} S.,  {Wandelt} B.,   {Wild} V.,  2013, \mn@doi
  [\mnras] {10.1093/mnras/stt523}, \href
  {https://ui.adsabs.harvard.edu/abs/2013MNRAS.432.2061C} {432, 2061}

\bibitem[\protect\citeauthoryear{{Clayton}, {Gordon}, {Bianchi}, {Massa},
  {Fitzpatrick}, {Bohlin}  \& {Wolff}}{{Clayton} et~al.}{2015}]{Claytonm31}
{Clayton} G.~C.,  {Gordon} K.~D.,  {Bianchi} L.~C.,  {Massa} D.~L.,
  {Fitzpatrick} E.~L.,  {Bohlin} R.~C.,   {Wolff} M.~J.,  2015, \mn@doi [\apj]
  {10.1088/0004-637X/815/1/14}, \href
  {https://ui.adsabs.harvard.edu/abs/2015ApJ...815...14C} {815, 14}

\bibitem[\protect\citeauthoryear{{Conroy}}{{Conroy}}{2013}]{conrey2013}
{Conroy} C.,  2013, \mn@doi [\araa] {10.1146/annurev-astro-082812-141017},
  \href {https://ui.adsabs.harvard.edu/abs/2013ARA&A..51..393C} {51, 393}

\bibitem[\protect\citeauthoryear{{Conroy}, {White}  \& {Gunn}}{{Conroy}
  et~al.}{2010a}]{Conroyb}
{Conroy} C.,  {White} M.,   {Gunn} J.~E.,  2010a, \mn@doi [\apj]
  {10.1088/0004-637X/708/1/58}, \href
  {https://ui.adsabs.harvard.edu/abs/2010ApJ...708...58C} {708, 58}

\bibitem[\protect\citeauthoryear{{Conroy}, {Schiminovich}  \&
  {Blanton}}{{Conroy} et~al.}{2010b}]{Conroy2010}
{Conroy} C.,  {Schiminovich} D.,   {Blanton} M.~R.,  2010b, \mn@doi [\apj]
  {10.1088/0004-637X/718/1/184}, \href
  {https://ui.adsabs.harvard.edu/abs/2010ApJ...718..184C} {718, 184}

\bibitem[\protect\citeauthoryear{{De Barros}, {Reddy}  \& {Shivaei}}{{De
  Barros} et~al.}{2016}]{DeBarros}
{De Barros} S.,  {Reddy} N.,   {Shivaei} I.,  2016, \mn@doi [\apj]
  {10.3847/0004-637X/820/2/96}, \href
  {https://ui.adsabs.harvard.edu/abs/2016ApJ...820...96D} {820, 96}

\bibitem[\protect\citeauthoryear{Draine}{Draine}{2003}]{Draine_2003}
Draine B.,  2003, \mn@doi [Annual Review of Astronomy and Astrophysics]
  {10.1146/annurev.astro.41.011802.094840}, 41, 241–289

\bibitem[\protect\citeauthoryear{{Draine} \& {Li}}{{Draine} \&
  {Li}}{2007}]{Draine&li}
{Draine} B.~T.,  {Li} A.,  2007, \mn@doi [\apj] {10.1086/511055}, \href
  {https://ui.adsabs.harvard.edu/abs/2007ApJ...657..810D} {657, 810}

\bibitem[\protect\citeauthoryear{{Fanelli}, {O'Connell}  \& {Thuan}}{{Fanelli}
  et~al.}{1988}]{Fanelli}
{Fanelli} M.~N.,  {O'Connell} R.~W.,   {Thuan} T.~X.,  1988, \mn@doi [\apj]
  {10.1086/166869}, \href
  {https://ui.adsabs.harvard.edu/abs/1988ApJ...334..665F} {334, 665}

\bibitem[\protect\citeauthoryear{{Fitzpatrick} \& {Massa}}{{Fitzpatrick} \&
  {Massa}}{1990}]{Fitzpatrick90}
{Fitzpatrick} E.~L.,  {Massa} D.,  1990, \mn@doi [\apjs] {10.1086/191413},
  \href {https://ui.adsabs.harvard.edu/abs/1990ApJS...72..163F} {72, 163}

\bibitem[\protect\citeauthoryear{{Fitzpatrick} \& {Massa}}{{Fitzpatrick} \&
  {Massa}}{2007}]{fitzpatric}
{Fitzpatrick} E.~L.,  {Massa} D.,  2007, \mn@doi [\apj] {10.1086/518158}, \href
  {https://ui.adsabs.harvard.edu/abs/2007ApJ...663..320F} {663, 320}

\bibitem[\protect\citeauthoryear{{F{\"o}rster Schreiber} et~al.,}{{F{\"o}rster
  Schreiber} et~al.}{2009}]{Forster}
{F{\"o}rster Schreiber} N.~M.,  et~al., 2009, \mn@doi [\apj]
  {10.1088/0004-637X/706/2/1364}, \href
  {https://ui.adsabs.harvard.edu/abs/2009ApJ...706.1364F} {706, 1364}

\bibitem[\protect\citeauthoryear{{Gordon}, {Clayton}, {Misselt}, {Land olt}  \&
  {Wolff}}{{Gordon} et~al.}{2003}]{Gordon03}
{Gordon} K.~D.,  {Clayton} G.~C.,  {Misselt} K.~A.,  {Land olt} A.~U.,
  {Wolff} M.~J.,  2003, \mn@doi [\apj] {10.1086/376774}, \href
  {https://ui.adsabs.harvard.edu/abs/2003ApJ...594..279G} {594, 279}

\bibitem[\protect\citeauthoryear{{Hao}, {Kennicutt}, {Johnson}, {Calzetti},
  {Dale}  \& {Moustakas}}{{Hao} et~al.}{2011}]{Hao2011}
{Hao} C.-N.,  {Kennicutt} R.~C.,  {Johnson} B.~D.,  {Calzetti} D.,  {Dale}
  D.~A.,   {Moustakas} J.,  2011, \mn@doi [\apj] {10.1088/0004-637X/741/2/124},
  \href {https://ui.adsabs.harvard.edu/abs/2011ApJ...741..124H} {741, 124}

\bibitem[\protect\citeauthoryear{{Hoang}, {Tram}, {Lee}  \& {Ahn}}{{Hoang}
  et~al.}{2019}]{Hoang}
{Hoang} T.,  {Tram} L.~N.,  {Lee} H.,   {Ahn} S.-H.,  2019, \mn@doi [Nature
  Astronomy] {10.1038/s41550-019-0763-6}, \href
  {https://ui.adsabs.harvard.edu/abs/2019NatAs...3..766H} {3, 766}

\bibitem[\protect\citeauthoryear{{Johnson} et~al.,}{{Johnson}
  et~al.}{2007}]{Johnson}
{Johnson} B.~D.,  et~al., 2007, \mn@doi [\apjs] {10.1086/522960}, \href
  {https://ui.adsabs.harvard.edu/abs/2007ApJS..173..392J} {173, 392}

\bibitem[\protect\citeauthoryear{{Kashino} et~al.,}{{Kashino}
  et~al.}{2013}]{Kashino}
{Kashino} D.,  et~al., 2013, \mn@doi [\apjl] {10.1088/2041-8205/777/1/L8},
  \href {https://ui.adsabs.harvard.edu/abs/2013ApJ...777L...8K} {777, L8}

\bibitem[\protect\citeauthoryear{{Kauffmann} et~al.,}{{Kauffmann}
  et~al.}{2003}]{Kauffmann_2003}
{Kauffmann} G.,  et~al., 2003, \mn@doi [\mnras]
  {10.1111/j.1365-2966.2003.07154.x}, \href
  {https://ui.adsabs.harvard.edu/abs/2003MNRAS.346.1055K} {346, 1055}

\bibitem[\protect\citeauthoryear{{Kennicutt} \& {Evans}}{{Kennicutt} \&
  {Evans}}{2012}]{Kennicutt2012}
{Kennicutt} R.~C.,  {Evans} N.~J.,  2012, \mn@doi [\araa]
  {10.1146/annurev-astro-081811-125610}, \href
  {https://ui.adsabs.harvard.edu/abs/2012ARA&A..50..531K} {50, 531}

\bibitem[\protect\citeauthoryear{{Kennicutt} Robert~C. et~al.,}{{Kennicutt}
  et~al.}{2009}]{Kennicutt2009}
{Kennicutt} Robert~C. J.,  et~al., 2009, \mn@doi [\apj]
  {10.1088/0004-637X/703/2/1672}, \href
  {https://ui.adsabs.harvard.edu/abs/2009ApJ...703.1672K} {703, 1672}

\bibitem[\protect\citeauthoryear{{Koyama}, {Shimakawa}, {Yamamura}, {Kodama}
  \& {Hayashi}}{{Koyama} et~al.}{2019}]{Koyama}
{Koyama} Y.,  {Shimakawa} R.,  {Yamamura} I.,  {Kodama} T.,   {Hayashi} M.,
  2019, \mn@doi [\pasj] {10.1093/pasj/psy113}, \href
  {https://ui.adsabs.harvard.edu/abs/2019PASJ...71....8K} {71, 8}

\bibitem[\protect\citeauthoryear{{Kreckel} et~al.,}{{Kreckel}
  et~al.}{2013}]{Kreckel}
{Kreckel} K.,  et~al., 2013, \mn@doi [\apj] {10.1088/0004-637X/771/1/62}, \href
  {https://ui.adsabs.harvard.edu/abs/2013ApJ...771...62K} {771, 62}

\bibitem[\protect\citeauthoryear{{Kriek} \& {Conroy}}{{Kriek} \&
  {Conroy}}{2013}]{Krieck&cor}
{Kriek} M.,  {Conroy} C.,  2013, \mn@doi [\apjl] {10.1088/2041-8205/775/1/L16},
  \href {https://ui.adsabs.harvard.edu/abs/2013ApJ...775L..16K} {775, L16}

\bibitem[\protect\citeauthoryear{{Kroupa}}{{Kroupa}}{2001}]{Kroupa_2001}
{Kroupa} P.,  2001, \mn@doi [\mnras] {10.1046/j.1365-8711.2001.04022.x}, \href
  {https://ui.adsabs.harvard.edu/abs/2001MNRAS.322..231K} {322, 231}

\bibitem[\protect\citeauthoryear{{Li} \& {Draine}}{{Li} \&
  {Draine}}{2001}]{Li2001}
{Li} A.,  {Draine} B.~T.,  2001, \mn@doi [\apj] {10.1086/323147}, \href
  {https://ui.adsabs.harvard.edu/abs/2001ApJ...554..778L} {554, 778}

\bibitem[\protect\citeauthoryear{{Liu} et~al.,}{{Liu} et~al.}{2013}]{Liu}
{Liu} G.,  et~al., 2013, \mn@doi [\apjl] {10.1088/2041-8205/778/2/L41}, \href
  {https://ui.adsabs.harvard.edu/abs/2013ApJ...778L..41L} {778, L41}

\bibitem[\protect\citeauthoryear{{Mart{\'\i}nez-Gonz{\'a}lez}, {W{\"u}nsch}  \&
  {Palou{\v{s}}}}{{Mart{\'\i}nez-Gonz{\'a}lez} et~al.}{2017}]{Martinez}
{Mart{\'\i}nez-Gonz{\'a}lez} S.,  {W{\"u}nsch} R.,   {Palou{\v{s}}} J.,  2017,
  \mn@doi [\apj] {10.3847/1538-4357/aa7510}, \href
  {https://ui.adsabs.harvard.edu/abs/2017ApJ...843...95M} {843, 95}

\bibitem[\protect\citeauthoryear{{Nandy}, {Thompson}, {Jamar}, {Monfils}  \&
  {Wilson}}{{Nandy} et~al.}{1975}]{Nandy75}
{Nandy} K.,  {Thompson} G.~I.,  {Jamar} C.,  {Monfils} A.,   {Wilson} R.,
  1975, \aap, \href {https://ui.adsabs.harvard.edu/abs/1975A&A....44..195N}
  {44, 195}

\bibitem[\protect\citeauthoryear{{Nandy}, {Morgan}, {Willis}, {Wilson},
  {Gondhalekar}  \& {Houziaux}}{{Nandy} et~al.}{1980}]{Nandy_80}
{Nandy} K.,  {Morgan} D.~H.,  {Willis} A.~J.,  {Wilson} R.,  {Gondhalekar}
  P.~M.,   {Houziaux} L.,  1980, \mn@doi [\nat] {10.1038/283725a0}, \href
  {https://ui.adsabs.harvard.edu/abs/1980Natur.283..725N} {283, 725}

\bibitem[\protect\citeauthoryear{{Narayanan}, {Conroy}, {Dav{\'e}}, {Johnson}
  \& {Popping}}{{Narayanan} et~al.}{2018}]{Narayanan}
{Narayanan} D.,  {Conroy} C.,  {Dav{\'e}} R.,  {Johnson} B.~D.,   {Popping} G.,
   2018, \mn@doi [\apj] {10.3847/1538-4357/aaed25}, \href
  {https://ui.adsabs.harvard.edu/abs/2018ApJ...869...70N} {869, 70}

\bibitem[\protect\citeauthoryear{{Osterbrock}}{{Osterbrock}}{1989}]{Osterbrock}
{Osterbrock} D.~E.,  1989, \skytel, \href
  {https://ui.adsabs.harvard.edu/abs/1989S&T....78..491O} {78, 491}

\bibitem[\protect\citeauthoryear{{Pei}}{{Pei}}{1992}]{Pei}
{Pei} Y.~C.,  1992, \mn@doi [\apj] {10.1086/171637}, \href
  {https://ui.adsabs.harvard.edu/abs/1992ApJ...395..130P} {395, 130}

\bibitem[\protect\citeauthoryear{{Prevot}, {Lequeux}, {Maurice}, {Prevot}  \&
  {Rocca-Volmerange}}{{Prevot} et~al.}{1984}]{Prevot}
{Prevot} M.~L.,  {Lequeux} J.,  {Maurice} E.,  {Prevot} L.,
  {Rocca-Volmerange} B.,  1984, \aap, \href
  {https://ui.adsabs.harvard.edu/abs/1984A&A...132..389P} {132, 389}

\bibitem[\protect\citeauthoryear{{Price} et~al.,}{{Price} et~al.}{2014}]{Price}
{Price} S.~H.,  et~al., 2014, \mn@doi [\apj] {10.1088/0004-637X/788/1/86},
  \href {https://ui.adsabs.harvard.edu/abs/2014ApJ...788...86P} {788, 86}

\bibitem[\protect\citeauthoryear{{Reddy} et~al.,}{{Reddy}
  et~al.}{2015}]{Reddy_2015}
{Reddy} N.~A.,  et~al., 2015, \mn@doi [\apj] {10.1088/0004-637X/806/2/259},
  \href {https://ui.adsabs.harvard.edu/abs/2015ApJ...806..259R} {806, 259}

\bibitem[\protect\citeauthoryear{Reddy et~al.,}{Reddy
  et~al.}{2020}]{Reddy_2020}
Reddy N.~A.,  et~al., 2020, \mn@doi [The Astrophysical Journal]
  {10.3847/1538-4357/abb674}, 902, 123

\bibitem[\protect\citeauthoryear{{Rocca-Volmerange}, {Prevot}, {Ferlet},
  {Lequeux}  \& {Prevot-Burnichon}}{{Rocca-Volmerange} et~al.}{1981}]{Rocco}
{Rocca-Volmerange} B.,  {Prevot} L.,  {Ferlet} R.,  {Lequeux} J.,
  {Prevot-Burnichon} M.~L.,  1981, \aap, \href
  {https://ui.adsabs.harvard.edu/abs/1981A&A....99L...5R} {99, L5}

\bibitem[\protect\citeauthoryear{{Salim} \& {Narayanan}}{{Salim} \&
  {Narayanan}}{2020}]{salimreview}
{Salim} S.,  {Narayanan} D.,  2020, \mn@doi [\araa]
  {10.1146/annurev-astro-032620-021933}, \href
  {https://ui.adsabs.harvard.edu/abs/2020ARA&A..58..529S} {58, 529}

\bibitem[\protect\citeauthoryear{{Salim}, {Boquien}  \& {Lee}}{{Salim}
  et~al.}{2018}]{Salim18}
{Salim} S.,  {Boquien} M.,   {Lee} J.~C.,  2018, \mn@doi [\apj]
  {10.3847/1538-4357/aabf3c}, \href
  {https://ui.adsabs.harvard.edu/abs/2018ApJ...859...11S} {859, 11}

\bibitem[\protect\citeauthoryear{{Salmon} et~al.,}{{Salmon}
  et~al.}{2016}]{Salmon}
{Salmon} B.,  et~al., 2016, \mn@doi [\apj] {10.3847/0004-637X/827/1/20}, \href
  {https://ui.adsabs.harvard.edu/abs/2016ApJ...827...20S} {827, 20}

\bibitem[\protect\citeauthoryear{{Seaton}}{{Seaton}}{1979}]{Seaton}
{Seaton} M.~J.,  1979, \mn@doi [\mnras] {10.1093/mnras/187.1.73P}, \href
  {https://ui.adsabs.harvard.edu/abs/1979MNRAS.187P..73S} {187, 73}

\bibitem[\protect\citeauthoryear{Seon \& Draine}{Seon \&
  Draine}{2016}]{Seon_2016}
Seon K.-I.,  Draine B.~T.,  2016, \mn@doi [The Astrophysical Journal]
  {10.3847/1538-4357/833/2/201}, 833, 201

\bibitem[\protect\citeauthoryear{{Shivaei} et~al.,}{{Shivaei}
  et~al.}{2018}]{Shivaei18}
{Shivaei} I.,  et~al., 2018, \mn@doi [\apj] {10.3847/1538-4357/aaad62}, \href
  {https://ui.adsabs.harvard.edu/abs/2018ApJ...855...42S} {855, 42}

\bibitem[\protect\citeauthoryear{{Shivaei} et~al.,}{{Shivaei}
  et~al.}{2020}]{Shivaei2020}
{Shivaei} I.,  et~al., 2020, \mn@doi [\apj] {10.3847/1538-4357/aba35e}, \href
  {https://ui.adsabs.harvard.edu/abs/2020ApJ...899..117S} {899, 117}

\bibitem[\protect\citeauthoryear{{Tremonti} et~al.,}{{Tremonti}
  et~al.}{2004}]{Tremonti2004}
{Tremonti} C.~A.,  et~al., 2004, \mn@doi [\apj] {10.1086/423264}, \href
  {https://ui.adsabs.harvard.edu/abs/2004ApJ...613..898T} {613, 898}

\bibitem[\protect\citeauthoryear{{Weingartner} \& {Draine}}{{Weingartner} \&
  {Draine}}{2001}]{Weingartner}
{Weingartner} J.~C.,  {Draine} B.~T.,  2001, \mn@doi [\apj] {10.1086/318651},
  \href {https://ui.adsabs.harvard.edu/abs/2001ApJ...548..296W} {548, 296}

\bibitem[\protect\citeauthoryear{{Wild} et~al.,}{{Wild} et~al.}{2011a}]{Wild_a}
{Wild} V.,  et~al., 2011a, \mn@doi [\mnras] {10.1111/j.1365-2966.2010.17536.x},
  \href {https://ui.adsabs.harvard.edu/abs/2011MNRAS.410.1593W} {410, 1593}

\bibitem[\protect\citeauthoryear{{Wild}, {Charlot}, {Brinchmann}, {Heckman},
  {Vince}, {Pacifici}  \& {Chevallard}}{{Wild} et~al.}{2011b}]{Wild11}
{Wild} V.,  {Charlot} S.,  {Brinchmann} J.,  {Heckman} T.,  {Vince} O.,
  {Pacifici} C.,   {Chevallard} J.,  2011b, \mn@doi [\mnras]
  {10.1111/j.1365-2966.2011.19367.x}, \href
  {https://ui.adsabs.harvard.edu/abs/2011MNRAS.417.1760W} {417, 1760}

\bibitem[\protect\citeauthoryear{{Witt} \& {Gordon}}{{Witt} \&
  {Gordon}}{2000}]{Witt}
{Witt} A.~N.,  {Gordon} K.~D.,  2000, \mn@doi [\apj] {10.1086/308197}, \href
  {https://ui.adsabs.harvard.edu/abs/2000ApJ...528..799W} {528, 799}

\bibitem[\protect\citeauthoryear{{Wuyts} et~al.,}{{Wuyts} et~al.}{2011}]{Wuyts}
{Wuyts} S.,  et~al., 2011, \mn@doi [\apj] {10.1088/0004-637X/738/1/106}, \href
  {https://ui.adsabs.harvard.edu/abs/2011ApJ...738..106W} {738, 106}

\bibitem[\protect\citeauthoryear{{Wuyts} et~al.,}{{Wuyts}
  et~al.}{2013}]{Wuyts_13}
{Wuyts} S.,  et~al., 2013, \mn@doi [\apj] {10.1088/0004-637X/779/2/135}, \href
  {https://ui.adsabs.harvard.edu/abs/2013ApJ...779..135W} {779, 135}

\bibitem[\protect\citeauthoryear{{Yoshikawa} et~al.,}{{Yoshikawa}
  et~al.}{2010}]{Yoshikawa}
{Yoshikawa} T.,  et~al., 2010, \mn@doi [\apj] {10.1088/0004-637X/718/1/112},
  \href {https://ui.adsabs.harvard.edu/abs/2010ApJ...718..112Y} {718, 112}

\bibitem[\protect\citeauthoryear{{Zeimann} et~al.,}{{Zeimann}
  et~al.}{2015}]{Zeimann}
{Zeimann} G.~R.,  et~al., 2015, \mn@doi [\apj] {10.1088/0004-637X/814/2/162},
  \href {https://ui.adsabs.harvard.edu/abs/2015ApJ...814..162Z} {814, 162}

\makeatother
\end{thebibliography}
\bsp	
\label{lastpage}
\end{document}